# The Causal Consciousness

Presynaptic β-neurexin promotes neuromediator release via vibrationally assisted multidimensional tunneling

**Danko Dimchev Georgiev, MD**

Medical University Of Varna, Varna 9000, Bulgaria
Department Of Emergency Medicine, Varna 9000, Bulgaria

Abstract

*Epiphenomenalism is shown to be absurd because the development of consciousness must be explainable through natural selection. A detailed neuromolecular basis of the neuromediator release is given and it is stressed on the possible key point where the quantum mind could act, namely presynaptic scaffold protein dynamics and detachment of the calcium sensor v-SNARE synaptotagmin-1. The β-neurexin molecules are tuned via fast propagating solitons by the quantum coherent microtubule network so that the β-neurexin molecule thermal vibrations could promote or suppress conformational changes via vibrational multidimensional tunneling, which drives synaptotagmin-1 detachment from the SNARE complex under calcium ion binding. Following the synaptotagmin-1 detachment membrane fusion takes place in SNARE dependent fashion and the presynaptic vesicle spills neuromediator in the synaptic cleft. Thus the quantum transfer of information causally affects the neuromediator release.*

Epiphenomenalism and causality

The *epiphenomenalism* is a sort of one-way dualism, in which *consciousness* is a product of brain processes but is itself without any *causal effect* on those processes. The central motivation for epiphenomenalism lies in the premise that all physical events have sufficient causes that lie within the class of physical events. If a mental event were something other than a physical event, then for it to make any causal contribution of its own in the physical world would require a violation of physical law.

Nevertheless the epiphenomenalism is absurd; it is just plain obvious that our pains, our thoughts, and our feelings make a difference to our (evidently physical) behavior; it is impossible to believe that all our behavior could be just as it is even if there were no pains, thoughts, or feelings (Taylor, 1963). The development of consciousness must be explainable through *natural selection*. But a property can be selected for only if it has an effect upon organisms' behavior. Therefore, consciousness (both qualia and intentional states) must have effects in behavior, i.e., epiphenomenalism is false. Today, this argument is generally associated with Popper & Eccles (1977). However it is an old argument and clear statements of it were offered by James (1879) and by Romanes

(1896). William James (1879, 1890) offered an intriguing variant of the argument from natural selection. If pleasures and pains have no effects, there would seem to be no reason why we might not abhor the feelings that are caused by activities essential to life, or enjoy the feelings produced by what is detrimental. Thus, if epiphenomenalism were true, the felicitous alignment that generally holds between affective valuation of our feelings and the utility of the activities that generally produce them would require a special explanation. Yet on epiphenomenalist assumptions, this alignment could not receive a genuine explanation. The felicitous alignment could not be selected for, because if affective valuation had no behavioral effects, misalignment of affective valuation with utility of the causes of the evaluated feelings could not have any behavioral effects either. Epiphenomenalists would simply have to accept a brute and unscientific view of pre-established harmony of affective valuation of feelings and the utility of their causes.

According to William S. Robinson (1999) James' argument can be met by supposing that both the pleasantness of pleasant feelings and the pleasant feelings themselves depend on neural causes. However these neural causes will give rise to something so useless (pleasantness of the pleasant feeling) as is the pleasant feeling itself (Georgiev, 2004). In its turn the epiphenomenalists must invoke another ad hoc supposition in order to explain why should the pleasantness of the pleasant feeling be tolerated by the natural selection! Thus we came to an infinite regress in order to keep fundamentally false theory. The conclusion is that the consciousness must be causally effective in order to fight for its existence!

## The mind-body interpretation of quantum mechanics

According to *orthodox scientific materialism*, *mental states* are *identical* with *brain states*; our thoughts and feelings, and our sense of self, are generated by electrochemical activity in the brain. This would mean either that one part of the brain activates another part, which then activates another part, etc., or that a particular region of the brain is activated spontaneously, without any cause, and it is hard to see how either alternative would provide a basis for a conscious self and free will. Francis Crick (1994) believes that consciousness is basically a pack of neurons. According to him the main seat of free will is probably in or near a part of the cerebral cortex known as the anterior cingulate sulcus, but he implies that our feeling of being free is largely, if not entirely, an illusion.

Those who reduce consciousness to a product of the brain disagree on the relevance of the quantum-mechanical aspects of neural networks: for example the late Roger Sperry (1994) and Daniel Dennett (1991) tend to ignore quantum physics, while Stuart Hameroff (1994) believes that consciousness arises from quantum coherence in microtubules within the brain's neurons.

In quantum mechanics with every elementary particle is associated a wave function from which we shall deduce the properties of the particle. The time evolution of the wave function is given by the Schrödinger equation. The wave function does not belong to physical reality (Nakhmanson, 2000; 2001). It is not in the real 3D-space, we cannot find and measure it there directly but still it determines the behavior of the quantum objects. The "waves of matter" are non-material. Einstein justly called them "Gespensterfelder".

Where is the ψ-function then? According to Raoul Nakhmanson (2001) the wave function is a mental construction in an abstract configuration space. Nevertheless it controls the behavior of material objects. How is it possible? We don't know how. But we know a really working example: The human consciousness can control behavior of material objects (though we also do not know how). Following this analogy Nakhmanson concludes that the particle has some kind of consciousness controlling its behavior, and the wave function is a product of this consciousness (or identical to the consciousness that enters the physical description of the world already at the fundamental level as irreducible ingredient). Von Neumann (1955), London & Bauer (1939), and Eugene Wigner (1961) suggested that human consciousness provokes a collapse of ψ-function. This would be possible only if the human consciousness is quantum phenomenon itself and is describable by common ψ-function of entangled protein states within the brain cortex! In this paper it will be described how quantum processes in the cytoskeleton (including neuronal microtubules) could be causally effective and thus major regulator of the neuromediator release in the synaptic clefts! Also this new form of *quantum dualism* bypasses the original epiphenomenalist's argument – that is consciousness could be *nonmaterial* but nevertheless it can control matter if it is a *quantum wave ψ*.

## Molecular steps in exocytosis

The targeting of transport vesicles to the correct membrane destination involves a much larger set of proteins than anticipated and several layers of protein-protein interactions. The process of 'vesicle targeting' includes all of the steps involved in delivering a newly formed transport vesicle to its target. In the broadest sense, targeting requires molecular motors and the actin and/or microtubule-based cytoskeletons to bring a vesicle from one part of the cell to another. Then tethering proteins collect and restrain vesicles at or near their cognate target membranes. Finally, a core layer of SNARE proteins interacts to bring vesicle membranes in exceedingly close apposition with their cognate target membranes, thereby driving membrane fusion. SNARE master proteins regulate each of these processes to enhance the spatial and temporal control of membrane-trafficking events.

*Vesicle trafficking.* Certain vesicle-trafficking steps require the translocation of a vesicle over a significant distance. For example, vesicles that carry proteins from the Golgi to the cell surface are likely to use motor proteins and a cytoskeletal track to get close to their target before tethering would be appropriate. Both the actin- and the microtubule-based cytoskeletons have been implicated in these processes, along with an increasing number of new motor proteins. Once the vesicles reach their targets, they come into contact with tethering factors that can restrain them there.

*Vesicle tethering.* According to Pfeffer (1999) it is useful to distinguish between the initial, loose 'tethering' of vesicles with their targets from the more stable, 'docking' interactions. Tethering could be considered to involve links that extend over distances of more than about half the diameter of a vesicle from a given membrane surface (>25 nm). Tethering interactions are likely to be involved in concentrating synaptic vesicles at the synapse (Robinson & Martin, 1998).

*Vesicle docking.* The term 'docking' is used to refer to the holding of two membranes within a bilayer's distance of one another (<5-10 nm). Stable docking probably represents several distinct, molecular states: the molecular interactions underlying the close and tight association of a vesicle with its target may include the molecular rearrangements needed to trigger bilayer fusion. A common feature of many proteins that function in vesicle tethering and docking is their propensity to form highly extended, coiled-coil structures (Pfeffer, 1996; Warren & Malhotra, 1998; Orci et al., 1998). Tethering and docking of a transport vesicle at the target membrane precedes the formation of a tight core SNARE complex.

Specific proteins called Rab GTPases regulate the tethering and the docking of the vesicles. Rab GTPases interconvert between inactive, GDP-bound forms and active, GTP-bound forms. GTP hydrolysis is not coupled to fusion; rather, Rab conformation, which depends on the guanine nucleotide to which a Rab is bound, regulates the recruitment of docking factors from the cytosol onto membranes. In this way, Rab GTPases regulate vesicle docking. To ensure that Rab proteins remain active on transport vesicles, the transport machinery may make use of a set of Rab-interacting proteins that lock the Rabs in their active conformations (Novick & Zerial, 1997; Schimmoller et al., 1998).

*Vesicle priming.* In neuronal exocytosis, the term 'priming' has been used to include all of the molecular rearrangements and ATP-dependent protein and lipid modifications that take place after initial docking of a synaptic vesicle but before exocytosis, such that the influx of calcium ions is all that is needed to trigger nearly instantaneous neurotransmitter release (Robinson & Martin, 1998; Rizo & Südhof, 1998). In other cell types, whose secretion is constitutive (i.e. continuous, calcium ion independent, non-triggered) there is no priming.

*Vesicle fusion.* The vesicle fusion is driven by SNARE proteins process of merging the vesicle membrane with the presynaptic one resulting in neuromediator release in the synaptic cleft. The fusion process is triggered by influx of calcium ions (Scheller, 1995; Südhof, 1995).

## Families of SNARE proteins

In general terms, SNAREs comprise distinct families of conserved membrane-associated proteins, which facilitate membrane fusion in eukaryotes. J. Rothman and colleagues first used the term SNARE to describe entities, which could participate in the binding of soluble factors (e.g. NSF - N-ethylmaleimide-sensitive fusion protein and its membrane-attachment proteins, SNAPs) to membranes derived from bovine brain homogenates (Söllner et al., 1993a; Rothman & Warren, 1994). These entities called SNAP receptors (hence the name, SNAREs) had been described previously as agents which could mediate membrane fusion events in an in vitro assay system, based upon protein transfer between donor and acceptor compartments. Two distinct categories of SNAREs have been described. SNAREs present on the vesicle (or donor) compartment are known as v-SNAREs, while those on target (or acceptor) compartment are known as t-SNAREs. The SNARE proteins are divided into three major families: the syntaxin, SNAP-25 and VAMP/synaptobrevin families (Gerst, 1999).

*Table 1 Families of v-SNAREs in mammals*

| v-SNAREs | Trafficking event /localization |
|---|---|
| VAMP1 / Synaptobrevin I | regulated exocytosis /synaptic vesicles |
| VAMP2 / Synaptobrevin II | regulated /synaptic vesicles, secretory granules |
| VAMP3 / Cellubrevin | regulated and constitutive /secretory granules, secretory vesicles |
| VAMP5 | constitutive /secretory vesicles, myotubes, tubulovesicular structures |
| VAMP7 / TI-VAMP | constitutive and regulated?/apical membrane, secretory granules, endosomes |
| VAMP8 / Endobrevin | endocytosis?/early endosomes |
| Synaptotagmin I, II III, V, X | regulated /synaptic vesicles, secretory granules |

*Table 2 Families of t-SNAREs in mammals*

| t-SNAREs | Trafficking event /localization |
|---|---|
| Syntaxin IA and IB | regulated /plasma membrane |
| Syntaxin 2 | constitutive /apical and basolateral membrane |
| Syntaxin 3 | regulated and constitutive?/apical membrane |
| Syntaxin 4 | regulated and constitutive?/basolateral membrane |
| SNAP25 A and B | regulated /plasma membrane |
| SNAP23 / Syndet | regulated and constitutive /plasma membrane |

## VAMP/synaptobrevin family

Synaptobrevin/VAMP (vesicle associated membrane protein) family members are small membrane proteins of about 120 amino acids. Structurally, they consist of a variable region of 25 -35 amino acids located at the amino terminus, followed by either one extended or two short amphipathic α-helical segments predicted to form coiled-coil structures, and have a transmembrane domain located at their carboxyl terminus. The putative α-helical regions, originally designated as Helix 1 (or H1) and Helix 2 (or H2) (Grote et al., 1995; Regazzi et al., 1996), are required for these SNAREs to mediate their protein-protein interactions, as shown by studies employing in vitro binding experiments (Hayashi et al., 1994), as well as in vivo exocytosis assays, using both yeast (Gerst, 1997) and mammalian cells (Regazzi et al., 1996). Targeting to synaptic-like vesicles is mediated by the H1, as shown using PC12 cells (Grote et al., 1995), whereas both helices participate in binding to the syntaxin and SNAP-25 t-SNAREs (Hao et al., 1997). Synaptobrevin/VAMP binding to these t-SNAREs in vitro has been shown to occur at regions of their carboxyl terminals, which have the potential to form coiled coils (Chapman et al., 1994; Hayashi et al., 1995; Kee et al., 1995).

## Syntaxin family

The first syntaxin family members were first identified as components of presynaptic nerve terminals (Bennett et al., 1992; Inoue et al., 1992) and only later as elements that participate in the binding of the SNAP proteins in vitro (Söllner et al., 1993a). Currently, there are 16 described members of the syntaxin family (excluding alternative spliceoforms) in mammalian cells, making it by far the largest SNARE family. The discrepancy between the number of syntaxin and VAMP or SNAP-25 isoforms is unclear, but does suggest some promiscuity in the protein-protein interactions of the latter two families. Currently, at least four syntaxins have been shown to localize to the plasma membrane and are thought to participate in exocytosis. These include syntaxins 1A, 1B and, possibly, 1C, as well as syntaxins 2 through 4 (Jagadish et al., 1997). Syntaxins 1A and B were the first syntaxin family members identified as SNAREs and are widely expressed in the nervous system (Bennett et al., 1993).

Syntaxins are membrane proteins, which possess a single transmembrane domain (Bennett et al., 1992; 1993). The carboxyl terminal H3 coiled-coil region (typically, residues 190-270) was found to mediate SNARE-SNARE interactions, binding to both synaptobrevin (Calakos et al., 1994) and SNAP-25 (Chapman et al., 1994; Kee et al., 1995). The H1 and H2 regions may inhibit the association of SNAREs with the H3 region, suggesting possible intramolecular regulation of SNARE assembly (Nicholson et al., 1998). Syntaxins possess a high degree of α-helicity present in the proposed SNARE binding domain - H3 (Fasshauer et al., 1997a; 1998; Zhong et al., 1997). Interactions with α-SNAP and, perhaps, other SNAP molecules are also mediated via this region (Hayashi et al., 1995; Hanson et al., 1995) and are strengthened by the core complex of syntaxin and synaptobrevin (McMahon & Südhof, 1995), and the ternary SNARE complex (Hayashi et al., 1995; McMahon & Südhof, 1995).

Syntaxins are involved in a number of other protein-protein interactions. Syntaxins have been shown to interact with voltage-gated calcium channel (VGCC) via the carboxyl coiled-coil domain, although residues embedded in the membrane may also participate in this interaction (el Far et al., 1995; Sheng et al., 1996). Electrophysiological studies in Xenopus oocytes revealed that full-length syntaxin stabilizes L-and N-type channels in an inactive state, whereas a carboxyl terminal truncated form was ineffective in doing so (Bezprozvanny et al., 1995). In addition, syntaxin 1A was also found to interact with, and to regulate, a nucleotide-gated chloride channel encoded by the cystic fibrosis gene (Naren et al., 1997). Thus, syntaxins appear to play an important role in channel conductance and influence their control over the fusion machinery.

Syntaxins were also found to bind synaptotagmin, a potential v-SNARE, in a calcium-dependent fashion (Kee et al., 1995; Chapman et al., 1995; Shao et al., 1997), as well as to a ubiquitously expressed form of SNAP-25, named SNAP-23/Syndet (Ravichandran et al., 1996; Wang, G. et al., 1997). The latter interaction implicates a role for syntaxins in constitutive exocytic events.

A function for a putative coiled-coil domain at the N-terminus of syntaxin may be to bind to nSec1 (Garcia et al., 1994; Pevsner et al., 1994), a regulator of SNARE-SNARE interactions. This interaction appears to be inhibited by phosphorylation of nSec1, suggesting a role for protein modification in directly regulating vesicle docking and fusion (Fujita et al., 1996). Another function for the N-terminus is to bind to the other C2 domain-containing proteins, which interact with calcium and phospholipids, such as Munc-13 (Betz et al., 1997) and syncollin (Edwardson et al., 1997).

## SNAP-25 family

The third SNARE family represented in the detergent-resistant 7S complex that binds α-SNAP and NSF in vitro (Söllner et al., 1993a; 1993b) is the synaptosomal protein, SNAP-25 (Oyler et al., 1989). SNAP-25 was originally identified as a synaptosomal protein of 25 kDa, which is expressed in nervous tissue (Oyler et al., 1989) and has been shown to exist in two variants: SNAP-25 A and B (Bark et al., 1995). SNAP-25 possesses heptad repeats that have a high probability of forming α-helices and coiled coils. A putative coiled-coil region located at the carboxyl terminus of SNAP-25 has been shown to mediate association with synaptobrevin/VAMP and syntaxin (Chapman et al., 1994). Despite the probability of forming coiled coils (based upon algorithmic analyses), the α-helical composition of recombinant SNAP-25 was found to be only marginal, as measured by circular dichroism (Fasshauer et al., 1997b; 1998). Yet α-helicity was found to increase dramatically upon binding to a soluble domain of syntaxin, indicating that the secondary (and tertiary) structures are induced by ternary complex formation (Fasshauer et al., 1997b; 1998).

SNAP-25, like synaptobrevin/VAMP and syntaxin, is sensitive to neurotoxins: BoNT/A and E (Blasi et al., 1993; Lawrence et al., 1996). Toxin treatment and cleavage, which result in truncation of the carboxyl terminus, weakens the interaction between SNAP-25 and VAMP in vitro (Chapman et al., 1994; Hayashi et al., 1995) and inhibits exocytosis from the nerve endings. SNAP-25 does not bear a membrane-spanning domain, although it is membrane-associated (Oyler et al., 1989). This association is likely to be mediated by palmitoylation, as mutants lacking in specific cysteine residues, which precede the coiled-coil domain, are found in the cytosolic fraction (Veit et al., 1996; Lane & Liu, 1997). The association of SNAP-25 with membranes may also be mediated by interactions with syntaxins, which complex together with SNAP-25 in axons (Garcia et al., 1995) and recycling synaptic vesicles (Walch-Solimena et al., 1995).

In addition to interactions with v-and t-SNAREs, SNAP-25 interacts with the proposed calcium sensor in exocytosis, synaptotagmin, even after toxin treatment (Schiavo et al., 1997). SNAP-25 has been suggested to bind to, and regulate, calcium channels (Rettig et al., 1996; Wiser et al., 1996; Yokoyama et al., 1997). These and earlier findings imply an additional regulatory role for SNAP-25 in the calcium-activated steps which lead to membrane fusion.

## Synaptotagmin family

Synaptotagmins were first identified as synaptic vesicle-associated membrane proteins that possess two calcium-binding motifs, called C2 domains (C2A and B), which are preceded by an α-helical coiled-coil domain (Perin et al., 1991a; 1991b). The C2A domain was shown to bind negatively charged phospholipids in a calcium-dependent fashion (Fukuda et al., 1996), whereas the C2B domain has been proposed to bind to inositol polyphosphates independent of calcium (Niinobe et al., 1994; Fukuda et al., 1994). More recently, the C2B domain has been shown to bind to phosphatidyl-inositol-3,4,5-triphosphate (PIP3 ) in the absence of calcium and PIP2 in its presence, suggesting that a lipid-interaction switch might occur during depolarization (Schiavo et al., 1996).

Interestingly, calcium binding also alters the protein-protein interactions of synaptotagmin. Calcium binding to the C2A domain increases the affinity of synaptotagmin for syntaxin (Chapman et al., 1995), whereas binding to the C2B domain appears to confer dimerization (Chapman et al., 1996; Sugita et al., 1996) and binding to the SV2 synaptic vesicle protein (Schivell et al., 1996). Synaptotagmin was also found to associate with voltage-gated calcium channels (VGCCs) (Petrenko et al., 1991; Leveque et al., 1994; Charvin et al., 1997; Sheng et al., 1997), like syntaxins, suggesting a similar role in channel regulation.

Synaptotagmin was found to bind stably to the SNAP-25 t-SNARE via the C2B domain, even after neurotoxin treatment (Schiavo et al., 1997), and was proposed therein to act as a possible calcium-regulated v-SNARE in exocytosis. The potential to bind calcium implicated synaptotagmin as a possible calcium sensor, which could confer the calcium-mediated signal in exocytic release (Brose et al., 1992; Elferink et al., 1993; Geppert et al., 1994a; Kelly, 1995). Studies using knockout mice revealed that homozygous knockouts died shortly after birth and had defects in evoked neurotransmitter release and synaptic transmission (Geppert et al., 1994a). This and some other works (Littleton et al., 1993; 1994; Broadie et al., 1994; Di Antonio & Schwarz, 1994) suggest that the evoked and spontaneous releases of neurotransmitter are likely to be mediated by separate pathways, one of which being devoid of synaptotagmin functioning.

SNARE masters and regulators

SNARE assembly culminating in membrane fusion is expected to occur only at the appropriate target membrane, in order to confer specificity to protein transport. Yet v-SNAREs and t-SNAREs involved in late trafficking events (i.e. post-Golgi) passage together through the secretory pathway and probably reside on identical ER-and Golgi-derived transport vesicles. Thus, some mechanisms must prevent nonproductive SNARE interactions from taking place earlier in the pathway. Likewise, other cellular mechanisms may mediate specific v-t-SNARE interactions in order to confer both temporal and spatial regulation of membrane docking and fusion. Thus, cells have evolved mechanisms to prevent nonproductive SNARE partnering early in the pathway and ready the SNAREs for assembly when they reach the appropriate target membrane. Therefore, in the hierarchy of events, which lead to membrane fusion, an activation step to remove these constraints might be a prerequisite for docking and fusion to proceed in vivo. This step might include dissociation of preformed SNARE complexes, as proposed by Ungermann et al. (1998), as well as the removal of other inhibitory constraints placed upon SNAREs. Such constraints include potential negatively acting SNARE regulatory proteins, which are designated as SNARE-masters (Gerst, 1998). By definition, these regulators are expected to bind directly to specific v-SNARE or t-SNARE partners and downregulate trafficking functions by modulating their entry into SNARE complexes. Thus, dissociation of SNARE regulators from SNAREs, or their inactivation, is expected to precede complex assembly and membrane fusion. Regulation of SNARE-master function also is an important step leading to exocytosis. This establishes a hierarchy in which a SNARE-master might find itself being more of a slave than an actual master.

n-Sec 1

It is the prototypic t-SNARE protector. This mammalian homologue of the yeast protein Sec1 binds directly and tightly to the t-SNARE on the presynaptic plasma membrane, syntaxin-1A; moreover, binding of n-Sec1 to syntaxin-1A blocks v-SNARE-t-SNARE association (Garcia et al., 1994). Sec1 homologues exist and could regulate other t-SNAREs. t-SNARE protectors could block v-SNARE binding by steric hindrance. Alternatively, as SNAREs are likely to exist in distinct conformations, SNARE protectors may bind preferentially to a particular conformation of a t-SNARE that interacts only weakly with a v-SNARE. Studies of the mammalian Sec1 homologs reveal that they are not components of the SNARE complex, but do bind directly to the syntaxin t-SNAREs (Garcia et al., 1994; Fujita et al., 1996; Hata et al., 1993a). Moreover, recombinant mammalian Sec1 was found to inhibit SNARE assembly from occurring in vitro and to dissociate from syntaxin after assembly had occurred (Garcia et al., 1994; Fujita et al., 1996; Hata et al., 1993a). This suggests that Sec1 homologs may restrict SNARE partnering by preventing the t-SNAREs (syntaxin and SNAP-25) from assembling into binary complexes. Thus, these proteins may exhibit negative regulatory functions.

The neuronal t-SNAREs, syntaxin and SNAP-25, are not restricted to sites of synaptic-vesicle fusion within the presynaptic membrane; they are also localized along the entire length of the axon, as well as in nerve terminals (Garcia et al., 1995). Thus, other proteins must specify vesicle-release sites and also regulate the SNAREs located elsewhere. However an unexpected link between n-Sec1 and proteins that connect presynaptic nerve terminals with their postsynaptic targets may hint at the spatial cues used for SNARE pairing at neurotransmitter-release sites in neurons (Butz et al., 1998). Specifically, n-Sec1 binds to a complex that includes Mint-1 (LIN-10), CASK (LIN-2) and the transmembrane protein β-neurexin; the extracellular portion of β-neurexin links to the postsynaptic-cell protein neuroligin to stabilize a functional synapse (Butz et al., 1998). The presynaptic components of this complex may act as a target specifier for vesicle release, and recruit SNAREs to this site through n-Sec1.

Interestingly, mammalian Sec1 is phosphorylated by protein kinase C in vitro (Fujita et al., 1996) and by Cdk5 (Shuang et al., 1998), a neuronal cyclin-dependent kinase. This modification was shown to inhibit interaction with syntaxin in vitro, suggesting a possible role for protein phosphorylation in the regulation of vesicle docking and fusion by Sec1 isoforms.

## Synaptophysins

A candidate SNARE-master for v-SNAREs involved in regulated exocytosis is synaptophysin, a membrane protein from synaptic vesicles (Jahn et al., 1985; Wiedenmann & Franke, 1985). Synaptophysin was found to complex directly with members of the synaptobrevin/VAMP family (Calakos & Scheller, 1994; Edelmann et al., 1995; Washbourne et al., 1995). Moreover, such complexes were devoid of the known t-SNARE partners, when antibodies were used to precipitate synaptophysin from detergent extracts (Edelmann et al., 1995; Washbourne et al., 1995; Galli et al., 1996). This suggests that the synaptophysin-synaptobrevin interaction prevents synaptobrevin from undergoing assembly with its partner t-SNAREs. However, the mechanism by which this protein disassociates to allow for formation of the fusion complex remains unknown.

## NSF/SNAP

A general class of SNARE regulators includes the NSF/Sec18 and SNAP/Sec17 protein families, which function upon membrane transport at various stages of the secretory pathway (Rothman & Orci, 1992; Rothman, 1994; Hay & Scheller, 1997). These proteins were originally proposed to act as agents which mediate the ATP-dependent step in membrane fusion (Rothman & Orci, 1992; Rothman, 1994; Wilson et al., 1989; Clary et al., 1990) and, later, to disassemble the ternary SNARE complex (Söllner et al., 1993a; Wilson et al., 1992). More recently, NSF/SNAP homologs in yeast have also been suggested to prime the docking step in SNARE assembly by dissociating preformed assemblages and allowing the syntaxin-like component to assume an 'activated' state (Ungermann et al., 1998). Based upon these and other studies, NSF is now beginning to look more like a molecular chaperone and SNARE regulator, rather than a component involved directly in the fusion event.

Recent structural analysis has demonstrated that NSF is a hexamer in solution and forms a hexagonal cylinder that is reminiscent of the ATPases known to act as molecular chaperones (Fleming et al., 1998). Thus, NSF regulation of the priming steps in vesicle docking may be based upon conferring a chaperone-like effect of protein unfolding and refolding in an ATP-dependent fashion, as proposed by Morgan & Burgoyne (1995). NSF/SNAP act together as regulators to ready SNAREs for forming associations in trans position (i.e. between donor and acceptor compartments). NSF/SNAP may act as general SNARE regulators to disassemble preformed v-t-SNARE complexes present in a cis conformation (that is SNARE assemblages formed in the same membrane) and later to regulate the association /dissociation of other, more specific, SNARE regulators, that is the SNARE-masters.

Rab GTPases

A large class of SNARE regulators is that of the rab family GTPases, which act upstream of SNARE complex assembly. Rabs have been suggested to regulate the specificity in membrane-trafficking steps due to their distinct subcellular localization and interaction with SNARE components (Pfeffer, 1996; Geppert & Südhof, 1998; Novick & Zerial, 1997; Bean & Scheller, 1997). Rab proteins regulate the fidelity of docking and fusion, and serve to dissociate proteins, which act as negative regulators for SNARE complex formation (i.e. SNARE-masters). In higher eukaryotes Rab3 GTPase, which is anchored to membranes by fatty acylation (Farnsworth et al., 1994), was shown early on to associate with secretory vesicles (von Mollard et al., 1990), although it can be removed from membranes in its inactive GDP-bound form by a soluble dissociation factor, GDI (Garrett et al., 1994). The fact that rab proteins associate with membranes in their GTP-bound state and are removed after GTP hydrolysis, suggests that they exist in a dynamic cycle of attachment and removal. This cycle has been proposed to be essential for the functioning of rabs in exocytosis and is supported by localization studies showing rab3 release from membranes after evoked stimulation (von Mollard et al., 1994; Jena et al., 1994).

Ironically, studies on rab3A ablation in mice revealed only minor behavioral changes, instead of catastrophic defects in synaptic transmission (Geppert et al., 1994b). This may result from compensation by other rab3 isoforms in nerve terminals, although some synapses were shown to be free of rab3. Despite this, an enhancement in evoked transmitter release was observed in certain synapses, resulting apparently from an increased number of vesicles undergoing fusion, as evidenced by paired-pulse experiments (Geppert et al., 1997).

The mechanism by which rabs function remains unclear, but appears likely to be at the level of vesicle docking, as evidenced from studies in yeast. Several putative effectors of the rab response have been identified, including rabphilin (Kishida et al., 1993; Shirataki et al., 1994; Li et al., 1994), RIM (Wang, Y. et al., 1997) and rabin (Brondyk et al.,

1995). Common to rabphilin and RIM are two calcium-dependent phospholipid-binding C2 domains and a zinc finger motif located at the amino terminus. The latter functions as the rab-binding motif (Li et al., 1994). Rabphilin is a vesicle-associated protein, which is recruited by rab3 and dissociates after GTP hydrolysis, and in the presence of GDI (Stahl et al., 1996). Microinjection of rabphilin or rabphilin fragments was shown to inhibit neurotransmitter release in squid giant axons and cortical granule release from mouse eggs (Masumoto et al., 1996; Burns et al., 1998), but both native and a membrane-anchored form of rabphilin induced insulin secretion from HIT-T15 cells. RIM, on the other hand, localizes to the presynaptic active zones in synapses, but like rabphilin associates only with the GTP-bound form of rab3 (Wang, Y. et al., 1997). Both proteins possess C2 domains that are likely to be relevant to phospholipid binding and to mediate calcium-dependent exocytosis. Thus, rabphilin and RIM and are proposed to act directly as rab3 effectors. Nevertheless, their direct targets and mechanism of action remain obscure.

## DOC2

These proteins are another family of C2 domain-containing proteins, like synaptotagmin and Munc-13, and as such are thought to play a role in calcium-regulated secretion. DOC2 was first isolated as protein expressed in nervous tissue, which localizes to synaptic vesicle-enriched fractions (Orita et al., 1995; Sakaguchi et al., 1995; Verhage et al., 1997). DOC2 binds to Munc13 (Orita et al., 1997) via phorbol ester binding to the C1 domain of Munc13, suggesting a dual role for diacylglycerol (DAG) regulation of both proteins. In addition, other work has shown that Munc18/nSec1 also binds to DOC2 and that the proteins are in competition for interactions with syntaxin (Verhage et al., 1997). Thus, DOC2 was proposed to act as a modulator of the syntaxin-Munc18 interaction and might function as part of an exchange mechanism to relieve the inhibition conferred upon the syntaxin t-SNARE by the Sec1 homolog (Verhage et al., 1997). That both Munc13 and Munc18/nSec1 interact with DOC2 suggests that a mutual relationship between these proteins is important for calcium-dependent exocytosis.

## Protein kinase C

Protein kinase C (PKC) is a calcium-dependent and diacylglycerol (DAG)-binding serine/threonine kinase that has been implicated in modulating synaptic transmission. This stems in part from studies, which show that neurotransmitter release from presynaptic terminals, and various other exocytic events are modulated by exposure to phorbol esters (Byrne & Kandel, 1996; Ktistakis, 1998), which are known PKC agonists. Moreover, PKC has been shown to phosphorylate substrates involved in vesicle fusion. PKC phosphorylates Munc18/nSec1, which inhibits the interaction with syntaxin in vitro (Fujita et al., 1996). PKC-mediated phosphorylation of SNAP-25 enhances catecholamine release from intact PC12 cells.

## Munc-13

Some phorbol ester-mediated effects upon membrane trafficking were shown to be PKC-independent, suggesting an alternative mechanism could be operant (Fabbri et al., 1994; Redman et al., 1997; Smith et al., 1998). Recently, Munc13 in mammals was shown to associate with the plasma membrane in a phorbol ester-dependent fashion and to modulate evoked and spontaneous neurotransmitter release in overexpression studies (Betz et al., 1998). Thus, unc-13 proteins are likely to modulate synaptic transmission by acting as a mediary of diacylglycerol-activated signaling events. Since unc-13 was also shown to bind directly to syntaxin (Betz et al., 1997) and DOC2 (Orita et al., 1997), these effects are likely to be elicited at the level of membrane fusion, although the exact mechanism has not yet been resolved.

## Syncollin

It is a small protein, which binds to the syntaxin t-SNARE at low levels of calcium (Edwardson et al., 1997). Addition of recombinant syncollin was also found to inhibit zymogen granule release in pancreatic cells, implying a possible regulatory role in exocytosis.

## Hrs-2

It is an ATPase, which binds to SNAP-25 in the absence of calcium and inhibits calcium-evoked release from PC12 cells (Bean et al., 1997).

## Complexins

Small soluble factors, called complexins, were found to compete with α-SNAP for binding to syntaxin and the core SNARE complex. Because they compete with SNAP, and not with synaptotagmin, and colocalize with the t-SNAREs, it has been suggested that they may regulate SNAP-SNARE interactions in a sequential fashion (McMahon et al., 1995).

## SV2

It is a synaptic vesicle-associated membrane protein (Feany et al., 1992) expressed as two isoforms, SV2A and SV2B (Bajjalieh et al., 1993), of which the SV2A form interacts with the C2B domain of synaptotagmin in the absence of calcium (Schivell et al., 1996). SV2 appears to be a transporter protein, due to its two sets of six membrane-spanning domains and homology to neurotransmitter transporters from presynaptic membranes (Feany et al., 1992). Like the possible regulation of VGCCs by syntaxin, SV2 could be also influenced by the actions of synaptotagmin and vice versa.

## Membrane lipids

The cytoplasmic leaflet of cellular membranes is characterized by acidic phospholipids, like phosphatidylserine and phosphoinositide (PI). Abundant evidence already implicates the phosphorylated products of PI in the regulation of membrane trafficking (De Camilli et al., 1996; Martin, 1997). Factors other than NSF were shown to be necessary for the ATP-dependent steps in calcium-stimulated exocytosis. These were shown to include phosphatidylinositol transfer protein (Sec14/PITP/PEP3/CAST) and phosphatidylinositol-4-phosphate 5-kinase (PIP5K) (Hay et al., 1995; Martin et al., 1995; Martin, 1997). These late ATP-requiring or 'priming' steps reveal an important role for 5-PIs in stimulus-coupled exocytosis (Hay et al., 1995; Banerjee et al., 1996). The nature of this requirement is not entirely clear, but is supported by studies, which show potential roles for PIP2 and PIP3 in membrane fusion. Antibodies directed against PIP2 were shown to inhibit calcium-stimulated release (Hay et al., 1995).

Schiavo et al., (1996) show that synaptotagmin binds to PIP3 in the absence of calcium and to PIP2 in its presence. These studies suggest that an alteration in lipid binding might occur during depolarization and might reflect changes in attachment sites during fusion. Other evidence strengthens the relationship between PI metabolism and exocytosis. For example, many exocytic and endocytic processes are sensitive to wortmannin, a specific inhibitor of PI3-kinase. Moreover, an inositol 5-phosphatase, synaptojanin, has been shown to be modified directly by phosphorylation and is required for synaptic vesicle endocytosis after neuromediator release (McPherson et al., 1996). Additionally, the AP-2 clathrin adaptor molecule is an inositol phosphate-and PI-binding protein, suggesting that PI modulates the assembly of clathrin coats (De Camilli et al., 1996; Gaidarov et al., 1996; Slepnev et al., 1998). Thus, involvement of these different PI-modifying and PI-binding factors in membrane trafficking suggests an important role for PI metabolism therein. While the precise roles are still not fully understood, the fact that the charge of PI can be modified dramatically by phosphorylation and dephosphorylation does make it an attractive candidate for a membranal component that can mediate protein interactions with bilayers, as well as influencing protein-protein interactions occurring within bilayers.

## The SNARE cycle

Membrane fusion in eukaryotes is controlled by a hierarchy of factors, which act upon SNAREs to regulate their assembly, leading to vesicle docking and fusion. SNAREs exist in several distinct states during the SNARE cycle, which leads to fusion.

*Phase 1.* SNAREs are initially found in a stable complex on the same membrane (i.e. arranged in cis position). This is likely to occur on SNAREs present both on vesicles and the plasma membrane. A cue, which is mediated presumably by vesicle encroachment to the acceptor membrane, allows NSF to bind ATP and associate (via SNAP) with ternary SNARE complexes, forming the 20S particle.

*Phase 2.* Subsequent hydrolysis of ATP by NSF results in dissociation of the particle and leaves SNAREs in an unfolded transitional state.

*Phase 3.* This now allows for the association with proteins that stabilize SNAREs in an uncomplexed (relative to other SNAREs) and activated state. These proteins include the SNARE-masters that are Sec1/Munc18 and synaptophysin. Prior to docking and fusion, the putative SNARE-masters (i.e.nSec1/Munc18 and synaptophysin) are bound to the syntaxin t-SNARE and synaptobrevin/VAMP v-SNARE, respectively.

*Phase 4.* Next, rab proteins in their activated GTP-bound state associate with SNAREs in their stable, SNARE-master-bound, uncomplexed state. This might occur directly or via their putative effectors (rabphilin, RIM etc.). Presumably, it occurs on both donor and acceptor membranes in order to the ready v-SNAREs and t-SNAREs on their respective membranes.

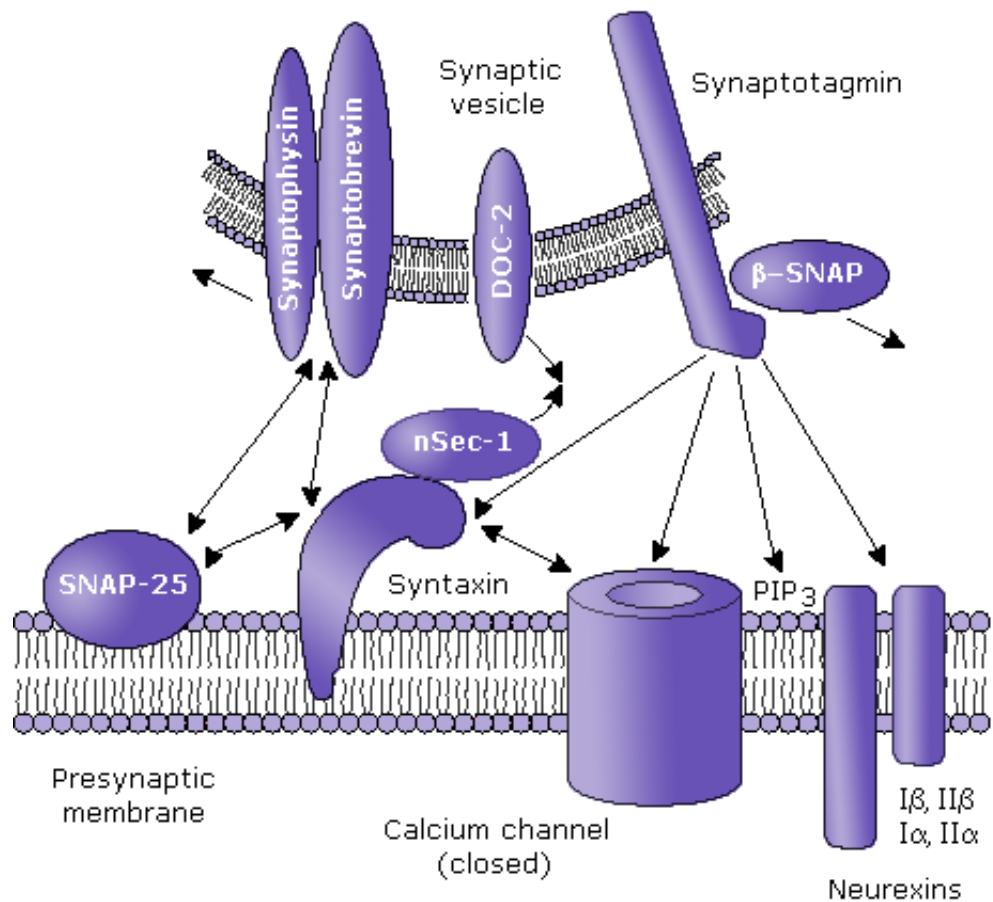

*FIG 1 The vesicle is ready for tethering and docking; arrows indicate some of the subsequent interactions. The activated rab proteins in their GTP-bound state and their effectors (e.g. rabphilin, RIM) are not shown!*

*Phase 5*. Vesicles may undergo docking and remain in the docked state. Hydrolysis of GTP by rab is expected to lead to dissociation of the SNARE-masters (i.e. nSec1 and synaptophysin), allowing for formation of a stable ternary SNARE complex between opposing membranes. Alternatively, GDP-GTP exchange mediated by rab GEFs might be sufficient to confer such an event. At this stage, synaptotagmin is likely to function as a component of the SNARE complex and act as a $Ca^{2+}$ - sensitive clamp. Synaptotagmin also interacts with the lipid molecules in the presynaptic membrane. Voltage-gated $Ca^{2+}$-channels (VGCCs) are expected to be associated with syntaxins at this stage, effectively maintaining the channel in its inactive state.

*Phase 6*. SNARE-mediated membrane fusion may occur immediately in constitutive secretory systems, but requires the influx of calcium in regulated secretory systems. Upon stimulus, nerve cell depolarization allows for the influx of calcium via VGCCs, resulting in the dissociation of synaptotagmin from the SNARE complex and allowing for membrane fusion to proceed in a SNARE-dependent fashion.

*Phase 7*. The membrane fusion proceeds in a SNARE-dependent fashion. Recent studies revealed that the v-and t-SNAREs are aligned in parallel within the core complex (Hanson et al., 1997; Lin & Scheller, 1997; Jahn & Hanson, 1998). Moreover, the individual SNAREs were found to undergo a significant increase in helicity upon association with their SNARE partners (Fasshauer et al., 1997a; 1997b), suggesting that formation of the complex is energetically favorable. This lends credence to the idea that SNARE assembly overcomes the electrostatic forces, which prevent membrane fusion from occurring. Studies on the structural arrangement of the exocytic SNARE complex have revealed that the α-helical regions of the v-and t-SNAREs assemble into a structure composed of four parallel helical domains (Hanson et al., 1997; Katz et al., 1998; Sutton et al., 1998). More important, crystallographic resolution of the complex, as determined by Brunger et al., reveals that it forms a tight four-helix bundle possessing both a packed charged and hydrophobic core, and grooved outer surface (Sutton et al., 1998). The α-helices are held together by hydrophobic interactions, as well as side chain-mediated hydrogen bonding and ionic interactions. Moreover, it was found that leucine zipper-like layers present within the complex stabilize the bundle and shield the ionic and hydrophobic core regions from interactions with the solvent (Sutton et al., 1998). The geometry of the helical domains of the fusion complex deviates from classical zippers, as well as from the standard coiled-coil domains. The shallow surface grooves, formed by helix association, are likely to allow for the binding of specific regulatory factors, such as SNAPs, NSF and so on, which have been proposed to mediate complex disassembly. The neuronal v- and t-SNAREs form exceedingly stable, parallel coiled-coil complexes that are stable in 0.1% SDS (Hayashi et al., 1994; Rizo & Südhof, 1998; Weis & Scheller, 1998; Hughson, 1998). The stability of this complex has led to the proposal that the energy gained from complex formation may be harnessed to drive the membrane-fusion reaction. Rothman and co-workers have shown that purified neuronal v-SNAREs and t-SNAREs, when reconstituted into distinct liposome vesicles, are themselves capable of driving liposome fusion, albeit at a rate that is significantly slower than the rate of exocytosis of synaptic vesicles. The slow rate observed could have been due, in part, to the presence of the syntaxin amino-terminal domain, which slows the assembly of the homologous yeast SNARE complexes as much as 2,000-fold in vitro (Nicholson et al., 1998). Release of this regulatory domain would be expected to enhance significantly the rate of tight SNARE pairing and membrane fusion. Thus, independent of any other proteins, paired SNAREs seem to be capable of stabilizing vesicles and target membranes in a closely apposed orientation, sufficient to drive liposome fusion (Weber et al., 1998).

*Phase 8*. Following fusion, SNAREs return to their stable cis-conformation. v-SNAREs are expected to cycle back to donor membranes via retrograde sorting mechanisms, whereas t-SNAREs may remain associated with the acceptor membrane.

## Activation barrier for neuromediator release

According to Beck & Eccles (1992) the synaptic exocytosis of neurotransmitters is the key regulator in the neuronal network of the neocortex. This is achieved by filtering incoming nerve impulses according to the excitatory or inhibitory status of the synapses. Findings by Jack et al. (1981) inevitably imply an *activation barrier*, which hinders vesicular docking, opening, and releasing of transmitter molecules at the presynaptic membrane upon excitation by an incoming nerve impulse. Redman (1990) demonstrated in single hippocampal pyramidal cells that the process of exocytosis occurs only with probability generally much smaller than one upon each incoming impulse. Deconvolution analysis by Sayer et al. (1989, 1990) has shown that an impulse invading bouton evokes exocytosis (release of neuromediator molecules) with probability 0.16-0.30.

There are principally two ways by which the barrier can be surpassed after excitation of the presynaptic neuron: the classical over-the-barrier thermal activation and quantum through-the-barrier tunneling. The characteristic difference between the two mechanisms is the strong temperature dependence of the former, while the pure quantum tunneling is independent of temperature, and only depends on the energies and barrier characteristics involved!

*Thermal activation.* This leads, according to Arrhenius' law, to a transfer rate, k, of

$$(1) \qquad k \sim V_C \exp\left(-\frac{E_A}{k_b T}\right),$$

where $V_C$ stands for the coupling across the barrier, and $E_A$ denotes the activation barrier.

*Quantum tunneling.* In this case with *pure tunneling* the transfer rate, k, is determined in a semiclassical approximation by

$$(2) \qquad k \approx \omega_0 \exp\left(-2\int_a^b \frac{\sqrt{2M[V(q)-E_0]}}{\hbar} dq\right)$$

with V(q) - the potential barrier; $E_0$ - the energy of the quasi-bound tunneling state; and $\omega_o = \dfrac{E_o}{\hbar}$ is the number of attempts that the particle undertakes to reach the barrier.

The quantum trigger model for exocytosis developed by Beck & Eccles (1992), Beck (1996) is based on the second possibility. The reason for this choice lies in the fact that thermal activation is a broadly uncontrolled process, depending mainly on the temperature of the surroundings, while quantum tunneling can be fine-tuned in a rather stable manner by adjusting the energy $E_0$ of the quasi-bound state or, equivalently, by regulating the barrier height (the role of the action potential).

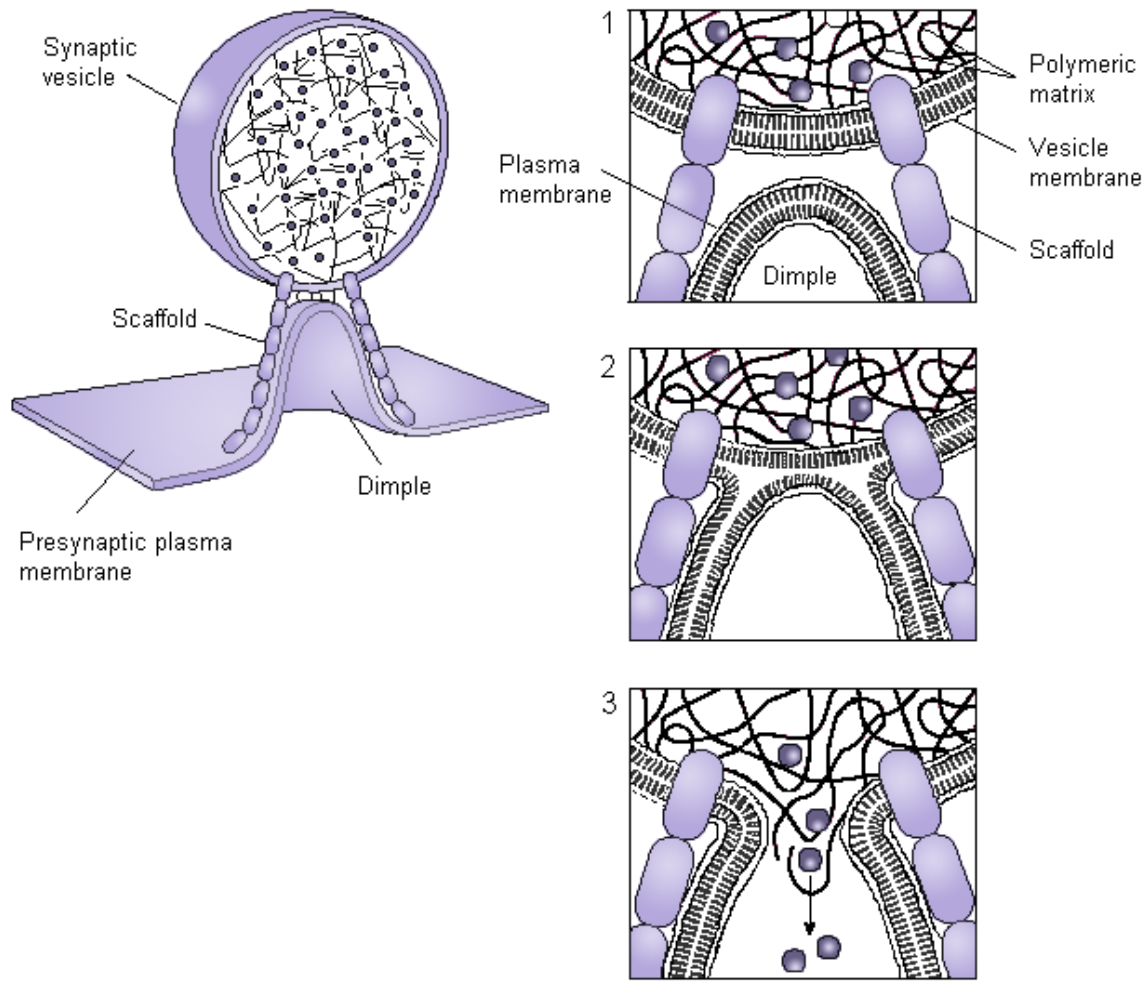

*FIG 2 Schematic representation of exocytosis (neuromediator release). The scaffold proteins drive vesicle and presynaptic membrane fusion (steps 1-3). When the fusion pore is dilated the polymeric matrix inside the vesicle undergoes phase transition and 'throws out' the neuromediator molecules (step 3).*

Causal model of Quantum mind action on exocytosis

One of the most interesting things about exocytosis is that the synaptic vesicles located in the *presynaptic vesicular grid* (hexagonal paracrystalline structure) of the axonal bouton manifest long-range correlations (Eccles, 1992; Beck & Eccles, 1992). There are about 40 synaptic vesicles in the vesicular grid but never does more than one vesicle emit transmitter molecules into the synaptic cleft after stimulation by a nerve impulse. This certainly means that that the vesicles in the vesicular grid do not act independently but rather that immediately after one vesicle is triggered for releasing its content the interaction between them blocks further exocytosis. The paracrystalline structure of the presynaptic vesicular grid makes it possible to have long-range interactions between the constituents, as is well known from ordered quantum systems. Beck & Eccles (1992) have calculated that the mass of the quasiparticle that promotes the exocytosis via tunneling is exactly that of a hydrogen atom. This supports the idea of vibrationally assisted tunneling as trigger for exocytosis as will be further revealed.

Considering the presented data it is safe to assume that only ¼ of the axonal terminals (boutons) release neuromediator under firing. This will produce complex geometric pattern of active (~25%) and silent (~75%) boutons that surely must be controlled by the conscious experience. The facts that a single axon forms synapses with dendrites from hundreds of other neurons and the relatively small number of released vesicles (only ¼ of the synapses release neuromediator) make the idea of the random (chaotic) vesicle release untenable i.e. it must be decided which postsynaptic neurons will be activated and which will not!

*The action potential.* Although the action potential at the axonal terminus (and the associated calcium ion entry) actually is needed for exocytosis, it may not be regulator of quantum consciousness action. In a previous paper (Georgiev, 2002) it was shown that quantum coherent cytoskeletons of different neurons could in principle be entangled via β-neurexin-neuroligin adhesion, which is connected to the cytoskeletons by protein lattice. The extracellular part of the adhesion is shielded against environmental decoherence through glycosaminoglycans interconnecting the presynaptic and postsynaptic membranes. The β-neurexin itself is supposed to affect exocytosis in a manner correlated with the quantum processes taking part in the neuronal cytoskeleton. *Such long-range correlations are specific characteristic for the quantum coherent systems!*

The β-neurexin was shown to be natural ligand for synaptotagmin, the calcium sensor in exocytosis. In the classical neuroscience after the ATP-dependent 'priming' all that is needed is calcium influx. Upon stimulus, nerve cell depolarization allows for the influx of calcium via VGCCs, resulting in the dissociation of synaptotagmin from the SNARE complex and allowing for membrane fusion to proceed in a SNARE-dependent fashion.

Although in vitro SNARE proteins themselves are capable of driving fusion of liposomes, in regulated exocytosis the calcium entry acts as a trigger mechanism. Because the triggering occurs only in ¼ of the cases it is essential to speculate that additional mechanism coupled to calcium ion influx takes part in cortical presynaptic terminals. The β-neurexin molecules could affect the key point namely the detachment of synaptotagmin-1 from the SNARE complex. If β-neurexin is capable to catalyze via quantum tunneling synaptotagmin-1 detachment from the SNARE complex under calcium ion entry the vesicle and synaptic membrane fusion will be successfully triggered. If so, it comes out that especially in the brain cortex calcium influx is not all that is needed! The process of synaptotagmin-1 detachment under calcium ion binding will have an energy barrier and although the barrier height could be dependent on membrane potential the process itself will still depend on β-neurexin tunnelling action. Thus exocytosis could be 'promoted' or 'locked' when needed.

*Soliton propagation and informational transfer.* The local membrane potential created electromagnetic field interacts with underlying microtubules generating fast propagating solitons that are capable of long-distance informational transfer. In order to causally determine exocytosis these solitons must transfer information faster than the membrane potential saltatory jumps in myelinated axons (v ~ 40-100 m/s). The needed time for quantum coherence (i.e. before collapse of the wave function to occur) must be comparable with the time of protein dynamics. Brunori et al. (1999) determine the protein dynamical time to be 10-15 ps; while Xie at al. (2000) have shown 15 ps long-lived quantum coherent amide I vibrational modes in myoglobin.

Caspi & Ben-Jacob (1999, 2002) show that the conformational changes in proteins and the associated protein folding could be induced by the propagation of a solitons along the molecular polypeptide chain, process known as a soliton mediated conformational transition (SMCT). They conclude that a soliton propagating along a protein backbone provides an effective mechanism for fast and deterministic change in the molecular conformation. Soliton creation could be induced spontaneously by thermal excitation, but there might exist other mechanisms for creation of solitons, such as interaction with "chaperones", enzymes that catalyze protein folding.

The interaction between the electromagnetic field and the neuronal microtubules that leads to sine-Gordon soliton propagation is described by Abdalla et al. (2001), which calculate the velocity of the soliton $v_0$ ~140 m/s. The frequency characteristic of their model is compatible with the transition period observed for the so-called conformational changes connected with tubulin dimer protein (namely 1-100 GHz).

We could therefore summarize the basic features of the proposed subneuronal quantum model in five crucial statements: (i) microtubules interact with the local electromagnetic field generating fast propagating solitons; (ii) solitons transfer quantum information; (iii) information can be stored in different information-holding configurations; (iv) the time for coherence must be comparable with time for protein dynamics i.e. 10-15 picoseconds; (v) the solitons affect both cytoskeletal and pre-synaptic scaffold protein dynamics.

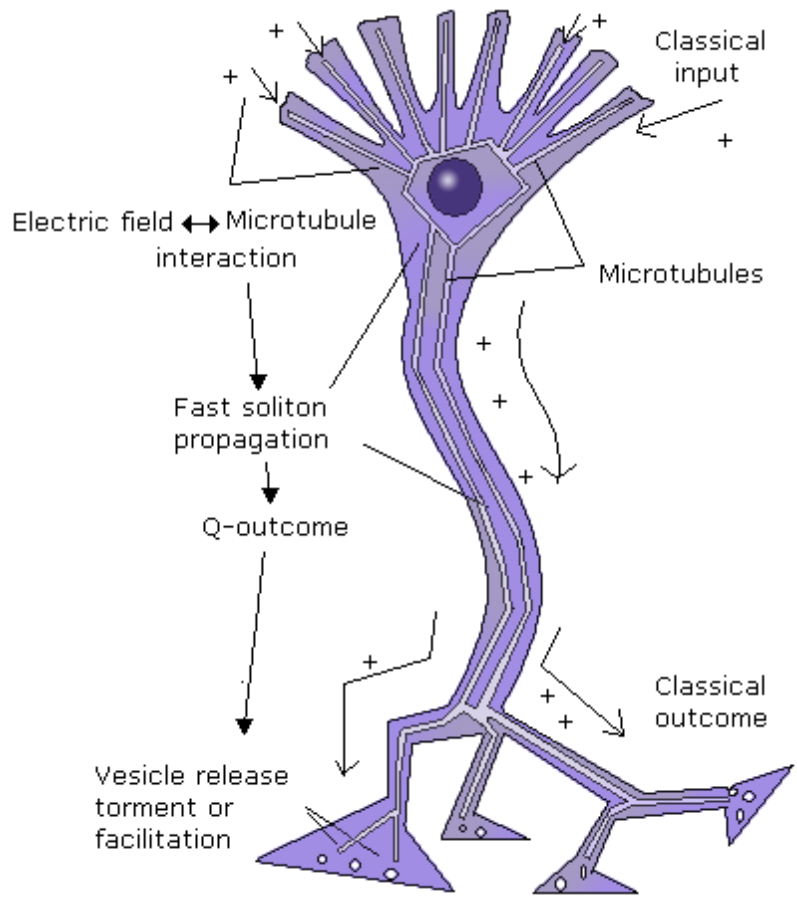

*FIG 3 Electromagnetic field interacts with microtubules generating fast solitons that propagate and transfer energy and information without dissipation. The solitons then could affect the presynaptic exocytotic machinery, so that upon membrane depolarization ($Ca^{2+}$ entry) to facilitate or torment vesicle fusion i.e. neuromediator release.*

*Probabilistic neuromediator release.* The collapse of the ψ-function is nonrandom nondeterministic (noncomputable) and thus it can be considered as determining the 'free will'. The probability of occurrence of natural phenomenon does not imply that this phenomenon is conscious. Instead if something is conscious you can only probably know (or guess) what is going next. The 'output' in the proposed β-neurexin promoted exocytosis is causally determined by the quantum events within the cortical cytoskeleton (including the microtubules) and thus is totally incompatible with the probabilistic exocytosis. The probability here comes from the fact that the consciousness describable by the ψ-function is unobservable!

*The temperature dependence.* While studying the enzyme action it was shown that the temperature dependence is not absolute criterion for quantum tunneling or not. The pure quantum tunneling reactions are temperature independent, because thermal activation of the substrate is not required to ascend the potential energy surface. Although the C-H and C-D cleavage by methylamine dehydrogenase required proton tunneling it was found to be strongly dependent on temperature, indicating that thermal activation or 'breathing' of the protein molecule is required for catalysis. Moreover, the temperature dependence of the reaction was found independent of isotope, reinforcing the idea that protein (and not substrate) dynamics drive the reaction and that tunneling is from the ground state. Good evidence is now available for *vibrationally assisted tunneling* (Bruno & Bialek, 1992; Basran et al., 1999) from studies of the effects of pressure on deuterium isotope effects in yeast alcohol dehydrogenase (Northrop & Cho, 2000). Combining the experimental evidence, the argument for vibrationally assisted tunneling is now compelling. The dynamic fluctuations in the protein molecule are likely to compress transiently the width of the potential energy barrier and equalize the vibrational energy levels on the reactant and product site of the barrier (Kohen & Klinman, 1999; Scrutton et al., 1999; Sutcliffe & Scrutton, 2000). Compression of the barrier reduces the tunneling distance (thus increasing the probability of the transfer), and equalization of vibrational energy states is a prerequisite for tunneling to proceed. Following transfer to the product side of the barrier, relaxation from the geometry required for tunneling "traps" the tunneled particle (H-nucleus) by preventing quantum 'leakage' to the reactant side of the barrier. Because the protein action in exocytosis includes also conformational changes (as well as tunneling of electrons) it can be considered that the quantum processes are dependent on temperature fluctuations but the energy gained from these fluctuations is used to "pump" biologically useful quantum transitions (Bruno & Bialek, 1992; Basran et al., 1999).

## Effects of vibrational excitation on multidimensional tunneling

Because the basic statement in the proposed causal model of quantum mind action is that β-neurexin promotes synaptotagmin-1 detachment from the SNARE complex under calcium ion binding via quantum tunneling (i.e. some kind of enzymatic action) or suppresses the detachment when needed, a specific mechanism for achieving this must be presented.

There exist two types of tunneling: one-dimensional and multi-dimensional. One of the most outstanding differences between one-dimensional and multidimensional tunneling is the existence of mixed tunneling. The mixed tunneling is such a tunneling that classical motion is allowed in one or more directions in the multidimensional space. In contrast in the pure tunneling classical motion is not allowed in any direction.

It was found that there are three kinds of vibrational modes with respect to the effects of the excitation on tunneling; (i) those which do not affect the tunneling, (ii) those which promote the tunneling, and (iii) those which suppress the tunneling. If there is no coupling between the tunneling coordinate and the co-ordinate transversal to it, the vibrational excitation in the latter does not affect energy splitting. This corresponds to the first type. When there is a coupling between the two coordinates, it is natural to expect from the analogy with the one-dimensional case that the vibrational excitations promote the tunneling. However, the experimental findings clearly show that the real proton tunneling is not so simple. This fact nicely exemplifies the complexity of multidimensional tunneling.

Takada & Nakamura (1995) show that depending on the topography of potential energy surface (PES) vibrational excitation either promote or suppress the tunneling and that the mixed tunneling plays an essential role in suppression and oscillation of energy splitting against vibrational excitation. The general Wentzel-Kramers-Brillouin (WKB) theory of multi-dimensional tunneling developed by Takada & Nakamura (1994, 1995) provides us with a clear conceptual understanding of the multidimensionality. The theory was formulated by solving the following basic problems: (i) construction of the semiclassical eigenfunction in classically allowed region according to the Maslov theory, (ii) its connection to the wave function in the classically inaccessible region, and (iii) propagation of the latter into the deep tunneling region.

It became clear that there exist two distinct tunneling regions: C-region where action is complex and I-region where action is pure imaginary. Tunneling in these regions is qualitatively quite different from each other; in the I-region the tunneling path can be defined by a certain classical trajectory on the inverted potential, while in the C-region there is no unique path and the Huygens type wave propagation should be applied.

The classical trajectories comprising a quantum eigenstate are confined within the distorted rectangular region (called R-region), although the much wider region is energetically allowed (see the oval region bounded by V=E in the next figure).

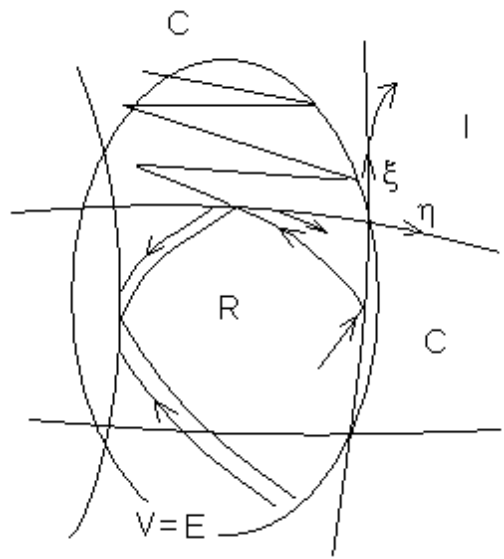

*FIG 4 Schematic drawing of the physical picture of tunneling obtained by the WKB theory.*

Tunneling proceeds first to the C-region where the motion in $\xi$-direction is non-classical (tunneling), while the motion in $\eta$-direction is still classical. Thus, Takada & Nakamura (1995) call this type of tunneling 'mixed tunneling.' At the boundary between the C-region and I-region, part of the tunneling wave enters into the I-region where no classical motion is allowed in any direction. This conventional type of tunneling is called

'pure tunneling.' One of the most outstanding differences between one-dimensional and multi-dimensional tunneling is the existence of mixed tunneling.

Table 3 summarizes the characteristics of the tunneling in each region. It should be noted that the wave function in the C-region has nodal lines, but that in the I-region does not. This plays an important role in the effects of vibrational excitation on tunneling.

*Table 3 Characteristics of multidimensional tunneling obtained by the WKB theory*

| Region | Characteristics |
|---|---|
| R-region | Classically allowed<br>Nodal pattern |
| C-region | Mixed tunneling<br>Tunneling in $\xi$-direction<br>Classical in $\eta$-direction<br>No tunneling path<br>Nodal lines |
| I-region | Pure tunneling<br>No classical motion<br>Tunneling path<br>Classical trajectory on $-V(x)$<br>No nodal line |

Takada & Nakamura (1995) have investigated the effects of vibrational excitation on tunneling using several model systems: tunneling energy splitting of excited states have been obtained by the exact quantum mechanical calculation and have been interpreted by the WKB theory and the sudden and adiabatic approximations. Various interesting effects have been found: 1) In the case of the symmetric mode coupling (SMC) model, vibrational excitation always promotes tunneling. 2) In the case of antisymmetric mode coupling (ASMC), pure tunneling in I-region is promoted by the excitation, while mixed tunneling in C-region is either promoted or suppressed. The latter is attributed to the phase cancellation in the overlap integral of Herring's formula. 3) In the case of squeezed potential, vibrational excitation suppresses the tunneling when the squeezing effect is sufficiently weak, while strong squeezing makes the tunneling irregular.

## Possible mechanism of β-neurexin action

Having presented the basis of multidimensional tunneling, we can further refine the proposed causal model of quantum consciousness claiming that the process of β-neurexin promoted synaptotagmin-1 detachment from the SNARE complex can be consequence of vibrational multidimensional tunneling. The different folding of the protein molecules leads to energy splitting - this means that between every two conformations there is an energy barrier. If we construct the potential energy surface (PES) then every stable conformation of the molecule will be at the bottom of a well (local minimum of the curve). The flipping between every two conformations must overwhelm the barrier (classically over-the-barrier, or via tunneling through-the-barrier). The tunneling could be mixed-multidimensional because the protein domain undergoing flipping into another conformation has also thermal fluctuation driven classical pathways.

We have already seen that synaptotagmin-1 binds $Ca^{2+}$ and membranes via its C2A-domain and plays an essential role in excitation-secretion coupling. Genetic studies have provided compelling evidence that synaptotagmin-1 function as the major $Ca^{2+}$-sensor in neuronal exocytosis. The ability of the C2A-domain of synaptotagmin to bind $Ca^{2+}$ ions and thus trigger exocytosis is strongly potentiated by anionic phospholipids (Brose et al., 1992). The interaction of C2A with membranes may directly participate in the $Ca^{2+}$-triggered lipid rearrangements that result in membrane fusion. However, synaptotagmin itself does not function as a fusogen (Brose et al., 1992) and spontaneous membrane fusion events do not require synaptotagmin (Littleton et al., 1994; Di Antonio & Schwarz, 1994).

Thomas & Elferink (1998) showed that a polylysine motif of synaptotagmin at residues 189-192 confers an inhibitory effect on secretion by recombinant synaptotagmin C2A fragments. The synaptotagmin C2A polylysine motif functions independently of calcium-mediated interactions with phospholipids and syntaxin-1A. Furthermore, α-latrotoxin reverses the inhibitory effect of injected recombinant C2A fragments, suggesting that they perturb the cellular calcium-sensing machinery by interfering with synaptotagmin activity in vivo. α-Latrotoxin is a potent neurotoxin from black widow spider venom (Rosenthal et al., 1990) proposed to trigger neurosecretion by binding to two types of cell surface receptors (neurexins and CIRL/latrophilin), which differ in their requirement for calcium binding (Hata et al., 1993; Krasnoperov et al., 1997; Lelianova, 1997). Whereas a role for neurexins in neurotransmitter release remains to be determined, CIRL and α-latrotoxin interactions in adrenal chromaffin cells seem to trigger secretion through the normal secretory apparatus. Thus, α-latrotoxin can bypass the calcium requirement for vesicle fusion and directly trigger neurotransmitter release.

According to Chapman & Davis (1998) the penetration of C2A domain of synaptotagmin into membranes may function to bring components of the fusion machinery into contact with the lipid bilayer to initiate exocytosis. $Ca^{2+}$ triggers the fusion of synaptic vesicles with the presynaptic plasma membrane on the submillisecond time scale, indicating that a limited number of conformational changes couple $Ca^{2+}$ to exocytotic membrane fusion. Using membrane-imbedded nitroxide quenchers, Chapman & Davis (1998) provided direct evidence for the partial insertion of a $Ca^{2+}$-binding "jaw" of C2A into the lipid bilayer. The $Ca^{2+}$- and lipid- binding sites overlap within C2A and support a $Ca^{2+}$-bridge model for the assembly of the $C2A/Ca^{2+}$-membrane complex (Bazzi & Nelsestuen, 1992). $Ca^{2+}$ makes direct contacts with both anionic head-groups of the lipids and with C2A. This model accounts for the lower affinity of specific C2A-domains for Ca2+ in the absence of lipids (Shao et al., 1996; Nalefski et al., 1997) since the $Ca^{2+}$-binding site may not be completely formed. It is supposed that the binding of metal ions to C2A-domain of synaptotagmin results in the extrusion of positively charged side chains (Shao et al., 1997). This conformational change results in the widening of the gap between the metal binding jaws such that it becomes large enough to accommodate a phospholipid head group. The extruded positively charged residues might mediate interactions with nearby anionic lipid head groups. This hypothesis is further supported by the high sensitivity of the C2A-domain-membrane interaction to increasing ionic strength and the presence of conserved basic residues in C2-domains that bind phosphatidylserine (Grobler & Hurley, 1997).

How might the $Ca^{2+}$-driven insertion of a $Ca^{2+}$-binding loop of C2A, into membranes, participate in excitation-secretion coupling? The $Ca^{2+}$-triggered penetration of C2A-loop 3 into membranes may function to destabilize the lipid bilayer by inducing lateral phase separations that facilitate bilayer fusion (Leckband et al., 1993). Crystallography studies indicate that portions of the C2 jaws of synaptotagmin are highly mobile in both the free and $Ca^{2+}$-bound states, so it is likely that while dipping into the intracellular leaflet of the presynaptic membrane the synaptotagmin molecule could be affected by its natural ligand β-neurexin. The β-neurexin molecule via its intracellular tail could catch synaptotagmin removing the synaptotagmin polylysine motif aside from interaction with the SNARE complex. The de-repressed SNARE complex then could make transition from the preformed 'loosely' associated state into 'tight' state that will subsequently drive membrane fusion (Stewart et al., 2000).

Burgoyne al. (2001) suppose *kiss-and-run* mechanism for neuromediator release at least in some cases. This means that the SNARE proteins open a pore in the docked vesicle, which when enough dilated can allow neuromediator release for a definite time. The pore could spontaneously close without membrane fusion to proceed. Instead the vesicle is swallowed and recycled without real endocytosis to take part! Wang, C. et al. (2001) investigated the synaptotagmin action on fusion pore kinetics and found that overexpression of synaptotagmin-1 prolonged the time from fusion time opening to dilation, whereas synaptotagmin IV shortened this time. It is interesting that in the proposed here model β-neurexin molecules catalyze detachment of synaptotagmin from the SNARE complex, de-repressing the fusion pore!

Fukuda et al., 2000 have shown that vesicle docking is mediated by WHXL motif in synaptotagmin molecule that binds to neurexins. So β-neurexin could indeed control the fusion events (i.e. fusion pore kinetics) in way causally determined by the microtubule quantum computation. This putative interaction is expected to take place in cortical neurons that have the potential to get entangled with other cortical neurons; i.e. to form the conscious 'hyperneuron'. The enormous diversity of neurexin subtypes expressed in the brain makes the proposition reasonable! Also the finding that α-latrotoxin could bypass the calcium dependent step in exocytosis (β-neurexin is also α-latrotoxin receptor) suggest that β-neurexin causal role in the neuromediator release is not unrealistic and that the proposed model is worth of further exploring.

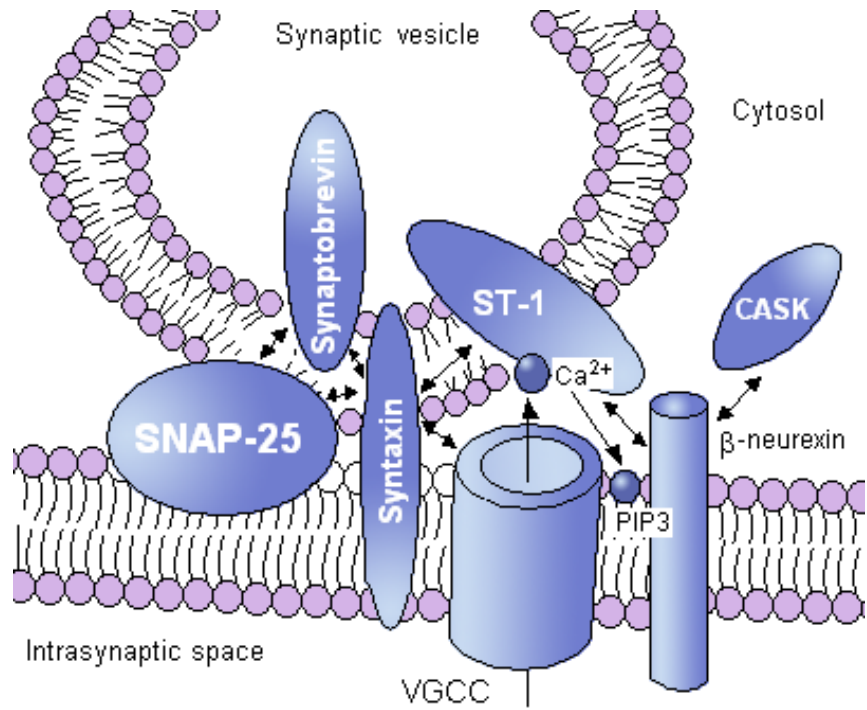

*FIG 5 Vibrationally assited tunneling in exocytosis: β-neurexin promotes synaptotagmin-1 detachment from the SNARE complex, de-repressing the fusion pore dilation. Legend: ST-1, synaptotagmin-1; PIP3, phosphatidyl inositol-3,4,5-triphosphate (ST-1-PIP3 interaction represented here is indeed the synaptotagmin 'jaw'-lipid binding); VGCC, voltage-gated calcium channel.*

It was already marked that the presynaptic scaffold proteins (CASK, Mint-1, Munc-18, syntaxin, n-Sec1) interact with proteins that connect presynaptic nerve terminals with their postsynaptic targets i.e. the β-neurexin-neuroligin-1 adhesion. Specifically, n-Sec1 binds to a complex that includes Mint-1 (LIN-10), CASK (LIN-2) and the transmembrane protein β-neurexin; the extracellular portion of β-neurexin links to the postsynaptic-cell protein neuroligin-1 to stabilize a functional synapse (Butz et al., 1998). Mint-1 also interacts with Munc-18 thus affecting syntaxin function. The presynaptic components of this complex may act as a target specifier for vesicle release, and recruit SNAREs to this site through n-Sec1. The β-neurexin molecules and the presynaptic scaffold proteins can be 'tuned' by the neuronal cytoskeleton via fast propagating solitons and their putative function is explainable in the frames of WKB theory. What deserves special attention is the *retrograde signaling at synapses* i.e. postsynaptic → presynaptic! At the present time there is not detailed investigation on β-neurexin/synaptotagmin interaction although the problem of neuronal exocytosis was rigorously exploited.

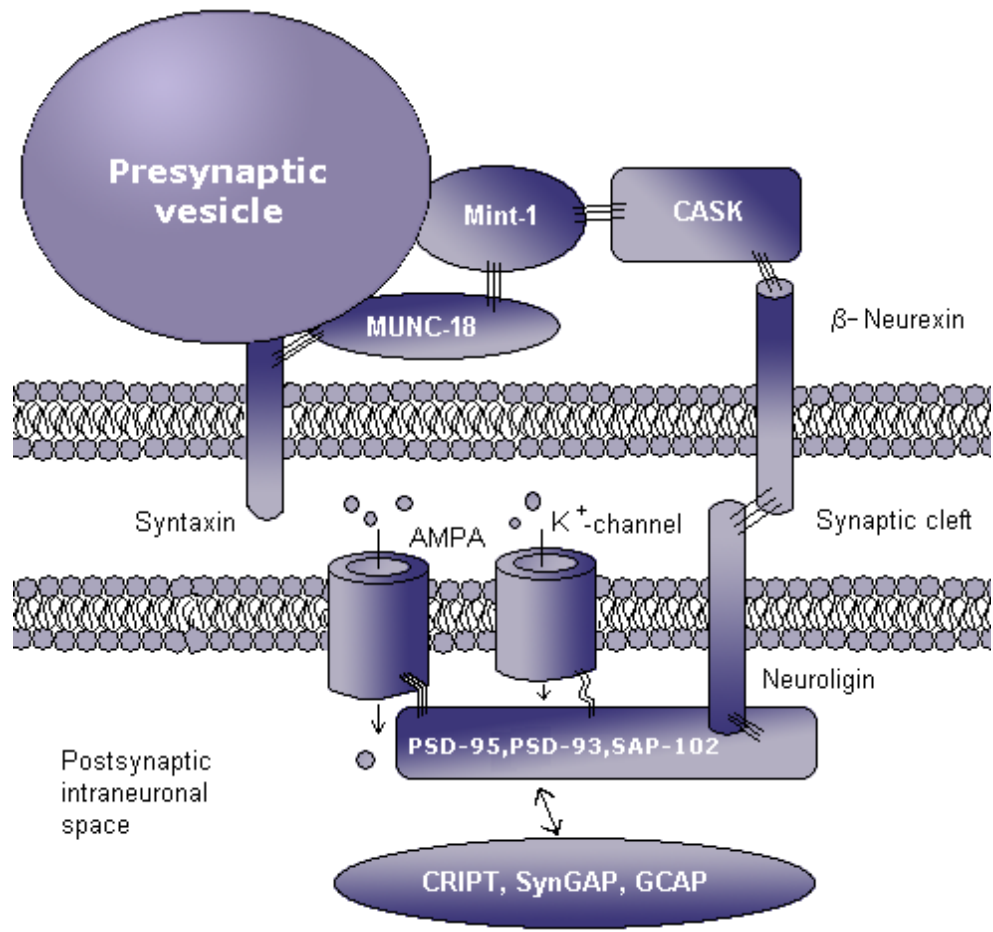

*FIG 6 The β-neurexin-neuroligin-1 junction is transferring information between the pre- and post-synaptic cytoskeletons. CASK and Mint 1 are presynaptic PDZ-domain proteins with a scaffold function; Munc18 and syntaxin are essential components of the presynaptic transmitter release machinery; PSD95, PSD93, SAP102, and S-SCAM are postsynaptic PDZ-domain proteins with a scaffold and assembly function that recruit ion channels (e.g. $K^+$-channels), neurotransmitter receptors (e.g. NMDA receptors) and other signal transduction proteins (GKAP, SynGAP, CRIPT).*